\documentclass[review]{article}
\usepackage{amsmath,amssymb}
\usepackage{amsthm}
\usepackage{titlesec}
\usepackage{indentfirst}

\usepackage{bbm}
\usepackage{mathrsfs}
\usepackage{amsfonts}
\usepackage{lineno,xcolor}

\usepackage{graphicx}

\theoremstyle{plain}
\newtheorem{thm}{Theorem}

\newtheorem{Lm}[thm]{Lemma}
\newtheorem{Df}[thm]{Definition}
\newtheorem{Rm}[thm]{Remark}

\begin{document}
\title{{\bf Lower bounds on the size of semi-quantum  automata}\thanks{This work is supported in
part by the National Natural Science Foundation of China (Nos. 61100001, 61472452, 61272058)}}

\author{ Lvzhou Li\thanks{Corresponding author.
  lilvzh@mail.sysu.edu.cn (L. Li)}, Daowen Qiu\thanks{issqdw@mail.sysu.edu.cn (D. Qiu)}\\ \small{Department of
Computer Science, Sun Yat-sen University, Guangzhou 510006, China}}
\date{ }
\maketitle

\begin{abstract} In the literature, there exist several interesting hybrid models of  finite automata
which have both quantum and classical states. We call them semi-quantum automata. In this paper, we compare the descriptional power of these models with that of DFA. Specifically, we  present a uniform method that gives a lower bound on the size of the three existing main models of semi-quantum automata, and this bound shows that semi-quantum automata can be at most exponentially more concise than DFA. Compared with a recent work (Bianchi, Mereghetti, Palano,  Theoret. Comput. Sci., 551(2014), 102-115), our method shows the following two advantages: (i) our method is much more concise; and (ii)  our method is universal, since it is applicable to the three existing main models of semi-quantum automata, instead of only a specific model.
\end{abstract}

\section{Introduction}

Quantum finite automata (QFA), as  theoretical models for quantum
computers with finite memory,  have been explored by many researchers.
  So far,  a variety of models of QFA have been introduced and explored to
various degrees (one can refer to a review article  \cite{QLMG12} and references therein).
Among these QFA,  there is a class of QFA that differ from others  by consisting of two interactive components: a quantum component and a classical one. We call them {\it semi-quantum automata} in this paper.
Examples of  semi-quantum automata are  {\it one-way QFA with control language} (CL-1QFA) \cite{Ber03}, {\it one-way QFA together with classical states} (1QFAC) \cite{QMS09}, and {\it one-way  finite automata with quantum and classical states} (1QCFA) \cite{ZQLG12}. Here ``one-way'' means that the automaton's tape head is required to move  right on scanning each tape cell.

These semi-quantum automata have been proved to not only recognize all regular languages,  but also  show superiority over DFA with respect to  descriptional power.
 For example, 1QCFA, CL-1QFA and 1QFAC were all shown to be much smaller than   DFA in accepting some languages (resolving some promise problems) \cite{GQZ14, MP06,QMS09,ZGQ14}. In addition,  a lower bound on the size of 1QFAC was given in \cite{QMS09}, which stated that  1QFAC can be at most exponentially more concise than DFA, and the bound was shown to be tight by giving some languages witnessing this exponential gap. Size lower bounds were also reported  for CL-1QFA in \cite{BMP14} and for 1QCFA in \cite{BMP14ciaa} (no detailed proof  was given in \cite{BMP14ciaa} for the bound of 1QCFA), but they were not proved to be tight.  By the way, we mention that the result obtained in \cite{BMP14ciaa} that 1QFCA recognize only regular languages follows directly from \cite{LF14}, although  a relatively complex procedure  was used in  \cite{BMP14ciaa} to deduce this result.

 Specially,  one can see that  complex technical treatments were used in \cite{BMP14} to derive the bound for CL-1QFA and  one may find that some key steps in \cite{BMP14} were confused  such that  the proof there may have some flaws,
 which will be explained more clearly in Section 4.  It is also worth mentioning that the method used in \cite{BMP14} is tailored for CL-1QFA and  is not easy to adopt to other models.

 Therefore, it is natural to ask: is there  a uniform and simple method giving lower bounds on the size of the above three semi-quantum automata? This is possible,  as 1QCFA, CL-1QFA and 1QFAC have the similar structure as shown in  \cite{LF14},  where they were described in a uniform way: a semi-quantum automaton can be seen as a two-component communication systems comprising a quantum component and a classical one, and they differ from each other mainly in the specific communication pattern: classical-quantum, or quantum-classical, or two-way.
It was  also proved in \cite{LF14} that the three models can be simulated by the model of QFA with mixed states and trace-preserving quantum operations(referred as MO-1gQFA) \cite{LQ12}.

 In this paper, by using the above result, we present a uniform method that gives a lower bound on the size of  1QCFA, CL-1QFA and 1QFAC, and this lower bound shows that they can be at most exponentially more concise than DFA. Specifically, we first obtain a lower bound on the size of MO-1gQFA and then apply it to the three hybrid models  by using the relationship between them and MO-1gQFA. Compared with a recent work \cite{BMP14}, our method is much more concise and universal, and it can be applied to the three existing main models of semi-quantum automata. In addition, our method may fix a potential mistake in \cite{BMP14} that will be indicated later on.

\section{Preliminaries}
Throughout this paper, for matrix (operator) $A$, $A^*$ and $A^\dagger$ denote the conjugate and conjugate-transpose  of $A$, respectively, and $\text{Tr}(A)$ and $rank(A)$  denote the
trace and rank of  $A$, respectively.
According to von Neumann's formalism of quantum mechanics, a quantum system is associated with a Hilbert space which is called the state space of the system. In this paper, we only consider finite dimensional spaces. A (mixed) state of a quantum system is represented by a density operator on its state space. Here a density operator $\rho$ on ${\cal H}$ is a positive semi-definite linear operator such that $\text{Tr}(\rho)= 1$. When $rank (\rho)=1$, that is, $\rho = |\psi\rangle\langle\psi|$ for some $|\psi\rangle\in {\cal H}$, then $\rho$ is called a pure state. Let $L({\cal H})$ and $D({\cal H})$ be the sets of linear operators and density operators on ${\cal H}$, respectively.

A trace-preserving quantum operation ${\cal E}$ on state space ${\cal H}$  is a linear map from $L(\mathcal{H})$ to itself that has an {\it operator-sum representation} as
\begin{align}
{\cal E}(\rho)=\sum_kE_k\rho E_k^\dagger,
\end{align}
with the  completeness condition $\sum_kE_k^\dagger E_k=I$, where $\{E_k\}$ are called operation elements of ${\cal E}$.

A general measurement is described by a collection $\{M_m\}$ of measurement operators,
where the index $m$ refers to the potential measurement outcome,  satisfying the  condition
$\sum_mM_m^\dagger M_m=I.$
If this measurement is performed on  a state $\rho$, then the classical outcome $m$ is obtained with the probability
$p(m)=\text{Tr}(M_m^\dagger M_m\rho)$, and the post-measurement state is
\begin{equation}\frac{M_m\rho M_m^\dagger}{\sqrt{p(m)}}.\end{equation}
For the case that $\rho$ is a pure state $|\psi\rangle$, that is, $\rho = |\psi\rangle\langle \psi|$, we have
$p(m)=\|M_m|\psi\rangle\|^2,$
and the state $|\psi\rangle$   ``collapses'' into the state
 \begin{equation}\frac{M_m|\psi\rangle}{\sqrt{p(m)}}.\end{equation}

A special case of general measurements is the projective measurement $\{P_m\}$ where $P_m$'s are orthogonal projectors.

$A\in L({\cal H})$ has the singular value decomposition \cite{HJ86,NC00} as follows:
\begin{align}
A=\sum_{i=1}^rs_i|u_i\rangle\langle v_i|, \label{SVD}
\end{align}
where $r=rank(A)$, $s_1,s_2,\dots,s_r\geq 0$  are called singular values of $A$, and
$\{|v_i\rangle\}_{i=1}^r, \{|u_i\rangle\}_{i=1}^r\subset {\cal H}$
are two orthonormal sets.

The  trace norm  of   $A\in L({\cal H})$ is defined as
$||A||_{tr}=\text{Tr}\sqrt{A^\dagger A}$.  By the  singular value decomposition in (\ref{SVD}), the trace norm can  be characterized by
 singular values  as \begin{align}||A||_{tr}=\sum_i
s_i.\end{align}
Note that if $A$ is positive semi-definite, then
$||A||_{tr}=\text{Tr}(A)$.

For $A, B\in L({\cal H})$, the trace distance between them is
\begin{align}D(A,B)=||A-B||_{tr}.\end{align}  The trace distance between two probability
distributions $\{p_x\}$ and $\{q_x\}$ is
\begin{align}
D(p_x,q_x)=\sum_x|p_x-q_x|.
\end{align}

Recall results about the trace distance from \cite{NC00} as follows.
\begin{Lm}[\cite{NC00}] Let $\rho$ and $\sigma$ be two density operators. Then we have
\begin{enumerate}
  \item [(i)] $D({\cal E}(\rho),{\cal E}(\sigma))\leq D(\rho,\sigma)$
 for any  trace-preserving quantum operation ${\cal E}$.
  \item [(ii)] $D(\rho,\sigma)=\max_{\{E_m\}}D(p_m,q_m)$ where
$p_m=\text{Tr}(\rho E_m)$, $q_m=\text{Tr}(\sigma E_m)$ and the
maximization is over all POVMs $\{E_m\}$.
\end{enumerate}\label{lm-distance}
\end{Lm}

A linear mapping $vec:  \mathbb{C}^{n\times n}\rightarrow \mathbb{C}^{n^2}$ which maps a $n\times n$ matrix to a $n^2$-dimensional column vector is defined as follows:
\begin{eqnarray}vec(A)((i-1)n+j)=A(i,j)\end{eqnarray}
 In other words, $vec(A)$ is the vector obtained by taking the rows of $A$, transposing them to form
column vectors, and stacking those column vectors on top of one another to form a single vector. For example, we have
\begin{eqnarray}A=\left(
                    \begin{array}{cc}
                      a & b \\
                      c & d \\
                    \end{array}
                  \right), &&
vec(A)=\left(
                    \begin{array}{cc}
                      a\\
                      b \\
                      c \\
                       d
                    \end{array}
                  \right).
\end{eqnarray}

If we let $|i\rangle$ be an $n$-dimensional column vector with the $i$th entry being 1 and else 0's, then $\{|i\rangle\langle j|: i,j=1,\cdots, n\}$ form a basis of  $\mathbb{C}^{n\times n}$. Therefore, the mapping $vec$ can also be defined as follows: \begin{eqnarray}vec(|i\rangle\langle j|)=|i\rangle|j\rangle.\end{eqnarray}
For any $|u\rangle, |v\rangle \in \mathbb{C}^{n}$, it is easy to verify \begin{align} vec(|u\rangle\langle v|)=|u\rangle (|v\rangle)^*.\end{align}

In this paper, the norm  of  $|v\rangle\in  \mathbb{C}^{n}$ is defined by $\||v\rangle\|=\sqrt{\Sigma_{i=1}^n |v_i|^2}$. For $A\in L({\cal H})$, we observe the following relation between the two norms $\|A\|_{tr}$ and $\|vec(A)\|$.

\begin{Lm} Let $A\in L({\cal H})$ and $dim({\cal H})=n$. Then we have
\begin{align}\|vec(A)\|\leq\|A\|_{tr}\leq\sqrt{n}\|vec(A)\|\end{align}\label{lm-norm}
\end{Lm}

{\noindent \it Proof. } Suppose $A$  has the singular value decomposition $A=\sum_{i=1}^rs_i|u_i\rangle\langle v_i|$. Then we have
\begin{align}vec(A)=\sum_{i=1}^rs_i|u_i\rangle(|v_i\rangle)^*.\end{align}
Thus we have
\begin{align}
\|vec(A)\|=\sqrt{\sum_{i=1}^r s_i^2}\leq \sum_{i=1}^r s_i=||A||_{tr}.
\end{align}
On the other hand, by the  Cauchy-Schwarz inequality we have
\begin{align}||A||_{tr}=\sum_{i=1}^r s_i\leq \sqrt{\sum_{i=1}^r s_i^2}\sqrt{\sum_{i=1}^r 1}\leq \sqrt{n}\|vec(A)\|.\end{align}

This completes the proof. \qed\\

\section{Definitions of automata}  
\begin{figure}[htbp]\centering
\includegraphics[width=60mm]{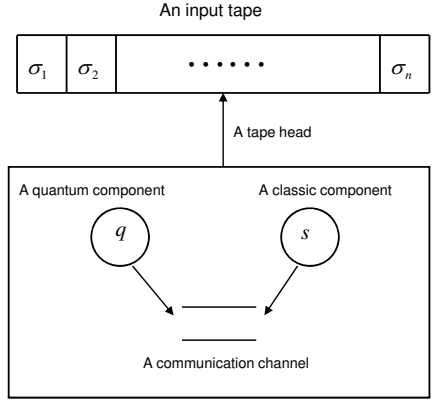}
\caption{ A diagram illustrating the idea behind  semi-quantum autoamta.  }\label{Model}
\end{figure}

In the literature, there exist some hybrid models of QFA that differ from other  QFA models  by consisting of two interactive components: a quantum component and a classical one. We call them {\it semi-quantum automata} in this paper.
As shown in \cite{LF14}, a semi-quantum automaton  can be depicted  in Fig. \ref{Model}, where an automaton comprises a quantum component, a classical component, a classical communication channel, and a classical tape head (that is, the tape head is regulated by the classical component). On scanning an input symbol, the quantum and classical components interact to evolve into new states, during which communication may occur between them.  In this paper, we focus on automata with a one-way tape head, that is,  after scanning an input symbol the model moves its tape head one cell right.

As shown in \cite{LF14},  there are three  models of semi-quantum automata fitting into  Fig. \ref{Model}, with the  essential difference being the specific communication pattern:
\begin{itemize}
  \item In CL-1QFA \cite{Ber03}, only quantum-classical communication is allowed, that is, the quantum component sends its measurement result to the classical component, but no reverse communication is permitted.
  \item In 1QFAC \cite{QMS09}, only classical-quantum communication is allowed, that is, the classical component sends its current state to the quantum component.
  \item In 1QCFA \cite{ZQLG12},  two-way communication is allowed: (1) first,  the classical component sends its current state to the quantum component; (2) second, the quantum component sends its measurement result to the classical component.
\end{itemize}

In the following, we recall the detailed definitions of the existing models of semi-quantum automata. One of such models is
 called {\it one-way QFA with control language} (CL-1QFA)\cite{Ber03},  defined as follows.
\begin{Df}
A  CL-1QFA  is a 7-tuple $${\cal
  A}=(Q, \Sigma, {\cal C},  q_1, \{U_\sigma\}_{\sigma\in\Sigma}, \mathcal{M}, {\cal L}),$$ where
 $Q$ is a finite set of quantum basis states, $\Sigma$ is a finite alphabet,
${\cal C}$ is a finite set of symbols  (measurement outcomes), $q_1\in Q$ is the initial quantum state, $U_{\sigma}$ is a unitary operator for each $\sigma\in\Sigma$,    $\mathcal{M}$ is a projective measurement given by a collection $\{P_c\}_{c\in{\cal C}}$ of projectors,
and ${\cal L}\subseteq{\cal C}^{*}$ is a regular language
(called a control language).\label{Df:CL-QFA}
\end{Df}

In CL-1QFA ${\cal A}$, on scanning  a symbol $\sigma$, a unitary operator $U_{\sigma}$ followed by the projective measurement $\mathcal{M}$ is performed on its current state. Thus, given an input string $x\in\Sigma^*$, the computation produces a
sequence $y\in {\cal C}^{*}$  of measurement results with a
certain probability $p(y|x)$ that is  given by
\begin{align}
p(y_1\dots y_{n}|x_1\dots
x_n)=\left\|\prod^{n}_{i=1}(P_{y_i}U_{x_i})|q_1\rangle\right\|^2,
\end{align}
 where we define  the ordered product $\prod_{i=1}^{n}A_i= A_nA_{n-1}\cdots A_1$. The input $x$ is said to be {\it accepted } if $y$ belongs to a fixed regular  language ${\cal
L}\subseteq {\cal C}^{*}$.
Thus the probability of ${\cal M}$
accepting $x$ is
\begin{align}
{\cal P}_{\cal A}(x)=\sum_{y_1\dots y_{n}\in {\cal L}}p(y_1\dots
y_{n}|x_1\dots x_n).\label{f_CL}
\end{align}

Recently, Qiu et al \cite{QMS09} proposed a new  model named {\it 1QFA together with classical states} (1QFAC), defined as follows.\footnote{In this paper we  consider only the case that 1QFAC are language acceptors, and one can refer to \cite{QMS09} for a more general definition.}

\begin{Df}
A 1QFAC ${\cal A}$ is defined by a 8-tuple
$${\cal A}=(Q, S, \Sigma, q_1, s_{1}, \{ U_{s,\sigma}\}_{s\in S, \sigma\in \Sigma} , \delta,
 \{ {\cal M}_s\}_{s\in S}),$$
where  $Q$ and $S$ are finite sets of quantum basis states and classical states, respectively, $\Sigma$  is a finite input alphabet,
$q_1\in Q$ and $s_1\in S$ are initial quantum and classical states, respectively,  $U_{s,\sigma}$ is a unitary operator on ${\cal H}_Q$ for each $s$ and $\sigma$,
  $\delta: S\times \Sigma\rightarrow S$ is a  classical transition  function, and for each $s$, ${
\cal M}_s$ is a projective measurement given by projectors $\{P_{s,a}, P_{s,r}\}$ where the two outcomes $a$ and $r$ denote acceptance and rejection, respectively.
\end{Df}

The machine starts with the initial states $s_{1}$ and $q_1$.
 On scanning an input symbol $\sigma\in\Sigma$,  $U_{s,\sigma}$ is first applied to the current quantum state, where $s$ is the current classical state; afterwards, the  classical state $s$ changes to $t=\delta(s,\sigma)$.
Finally, when the whole input string is finished,    a measurement  ${\cal M}_s$ determined by the last classical state is performed on the last quantum state, and the input is accepted if the outcome $a$ is observed.  Therefore, the probability of 1QFAC ${\cal A}$ accepting $x=x_1x_2\cdots x_n\in\Sigma^*$ is given by
\begin{align}{\cal P}_{\cal A}(x)=\|P_{s_{n+1}, a}U_{s_n,x_n}\cdots U_{s_2,x_2}U_{s_1,x_1}|q_1\rangle\|^2 \label{prob:1qfac}\end{align}
where $s_{i+1}=\delta(s_i,x_i)$ for $i=1,\cdots, n$.

Ambainis and Watrous \cite{AW02} proposed the model of {\it two-way QFA with quantum and classical states} (2QCFA).  As proved in \cite{AW02}, 2QCFA  can recognize
non-regular language $L_{eq}=\{a^{n}b^{n}|n>0\}$ in polynomial time and  the palindrome language $L_{pal}=\{x\in
\{a,b\}^{*}|x=x^{R}\}$ in exponential time, which shows the superiority of 2QCFA over their classical counterparts.   In the following we  recall 1QCFA \cite{ZQLG12}, a one-way variant of 2QCFA. Note that in this paper the notion of 1QCFA  is  slightly more general than the one  in  \cite{ZQLG12}.
The reason for why we adopt the current definition is that it has a more succinct form  which  simplifies some notations (for example, in our version we need give only the set of general measurements, instead of two sets: unitary operators and projective measurements). It is, however, worthwhile  to emphasize that all  results obtained in this paper  hold surely  for the model in \cite{ZQLG12}.
\begin{Df}
A 1QCFA  is specified by a 9-tuple
$$
\mathcal{A}=(Q,S,\Sigma, {\cal C}, q_1, s_{1}, \{\Theta_{s,\sigma}\}_{s\in S,\sigma\in\Sigma},\delta,S_a)
,$$
where
$Q$ and  $S$  are finite sets of quantum and classical states, respectively, $\Sigma$ is a finite input alphabet,  ${\cal C}$ is a finite set of symbols (measurement outcomes), $q_1\in Q$ and  $s_1\in S$  are initial quantum and classical states, respectively,     $\Theta_{s,\sigma}$  for each $s$ and $\sigma$ is a  general measurement  on ${\cal H}_Q$ with outcome set ${\cal C}$, $\delta: S\times \Sigma\times C\rightarrow S$ specifies the classical state transition, and $S_a\subseteq S$ denotes a set of accepting states.\label{Df-1QCFA}
\end{Df}

On scanning a symbol $\sigma$, at first the general measurement $\Theta_{s,\sigma}$, determined by the current classical state $s$ and the scanned symbol $\sigma$, is performed on the current quantum state,  producing some outcome $c\in {\cal C}$; then the classical
 state  $s$ changes to $s'=\delta(s, \sigma, c)$ by reading $\sigma$ and $c$. After scanning all input symbols, ${\cal A}$ checks whether its classical state is in $S_a$. If yes,  the input is accepted; otherwise, rejected.  Therefore, the probability of 1QCFA ${\cal A}$ accepting $x=x_1x_2\cdots x_n\in\Sigma^*$ is given by
\begin{align}{\cal P}_{\cal A}(x)=\sum_{c_1c_2\cdots c_n\in{\cal C}^n}\chi_a(s_{n+1})\|M^{c_n}_{s_n,x_n}\cdots M^{c_2}_{s_2,x_2}M^{c_1}_{s_1,x_1}|q_1\rangle\|^2 \label{prob:1qfac}\end{align}
where:
\begin{itemize}
  \item [(1)] $\chi_a: S\rightarrow\{0,1\}$ is defined by  $\chi_a(s)=\left\{
                                                                \begin{array}{ll}
                                                                  1, & \hbox{if $s\in S_a$;} \\
                                                                  0, & \hbox{otherwise.}
                                                                \end{array}
                                                              \right.
  $
  \item  [(2)] $\{M^{c}_{s,\sigma}\}_{c\in {\cal C}}$ are measurement operators of $\Theta_{s, \sigma}$.
  \item  [(3)] $s_{i+1}=\delta(s_i,  x_i, c_i)$ for $i=1,\cdots, n$.
\end{itemize}

\begin{Rm}In Fig.\ref{Model}, let $Q$ be the set of basis states of the quantum component and $S$ be the set of states of the classical component. Let $q=|Q|$ and $k=|S|$. Then we say that the semi-quantum automaton has $q$ quantum basisi states and $k$ classical states. For CL-1QFA, $S$ denotes the  set of states of the minimal DFA accepting the control language $L$.
\end{Rm}

Recall the model of MO-1gQFA  which has mixed states and trace-preserving quantum operation as follows \cite{LQ12}.

\begin{Df}
An MO-1gQFA ${\cal A}$ is a five-tuple ${\cal A}=\{ {\cal
H},\Sigma,\rho_0,\{{\cal E}_\sigma\}_{\sigma\in\Sigma},
P_{acc}\}$, where ${\cal H}$ is a finite-dimensional Hilbert
space, $\Sigma$ is a finite input alphabet, $\rho_0$, the initial
state of ${\cal A}$, is a density operator on ${\cal H}$, ${\cal
E}_\sigma$ corresponding to $\sigma\in\Sigma$ is a
trace-preserving quantum operation acting on ${\cal H}$, $P_{acc}$
is a projector on the subspace called accepting subspace of ${\cal
H}$. Denote $P_{rej}=I-P_{acc}$, then $\{P_{acc}, P_{rej}\}$ form
a projective measurement on ${\cal H}$. Let $dim ({\cal H})=n$. Then we call ${\cal A}$ is an $n$-dimensional MO-1gQFA.
\end{Df}

On the input word $x_1x_2\dots x_n\in\Sigma^{*}$, the
above MO-1gQFA ${\cal A}$ proceeds as follows: the quantum
operations ${\cal E}_{x_1},{\cal E}_{x_2},\dots,{\cal
E}_{x_n}$ are performed on $\rho_0$ in succession, and then
the projective measurement $\{P_{acc}, P_{rej}\}$ is performed on
the final state, obtaining the accepting result with a certain
probability. Thus,  MO-1gQFA  ${\cal A}$ defined above induces a
function ${\cal P}_{\cal A}: \Sigma^*\rightarrow [0,1]$ as
\begin{align}
{\cal P}_{\cal A}(x_1x_2\dots x_n)=\text{Tr}(P_{acc}{\cal
E}_{x_n}\circ\dots\circ{\cal E}_{x_2}\circ{\cal
E}_{x_1}(\rho_0)),
\end{align}
where ${\cal E}_2\circ{\cal E}_1(\rho)$ stands for  ${\cal
E}_2({\cal E}_1(\rho))$.  In fact, for every $x\in\Sigma^*$,
${\cal P}_{\cal A}(x)$ represents the probability that ${\cal A}$ accepts
$x$.

A DFA is a five-tuple ${\cal A}=(S,\Sigma,s_1,\delta,S_a)$ where $S$ is a finite state set, $\Sigma$ is a finite alphabet, $s_1\in S$ is the initial state, $S_a\subseteq S$ is the set of accepting states, and $\delta: S\times\Sigma\rightarrow S $ is the transition function: $\delta(s,\sigma)=t$ means that the current state $s$  changes to $t$ when scanning $\sigma$. Furthermore, $\delta$ can be extended to $\delta^*: S\times\Sigma^*\rightarrow S$ by defining:
 i) $\delta^*(s,\epsilon)=s$, and ii) $\delta^*(s,x\sigma)=\delta(\delta^*(s,x),\sigma)$ where $x\in\Sigma^*$ and $\sigma\in \Sigma$. ${\cal A}$ is said to accept $x\in\Sigma^*$, if $\delta^*(s_1,x)\in S_a$.

\section{The main results}

Some recent work (e. g. \cite{GQZ14,MP06,QMS09, ZGQ14}) shows that semi-quantum automata like CL-1QFA, 1QFAC and 1QFAC  are much more concise than equivalent DFA. In order to see the limit of the descriptional power of these models, size lower bounds were given in \cite{BMP14,BMP14ciaa,QMS09} for these models.

 However, it is worth mentioning that there were some potential flaws in the procedure to obtain lower bounds for CL-1QFA in \cite{BMP14}. Indeed, one may find that in Lemma 5 of \cite{BMP14} the two states $\varphi$ and $\varphi'$ were required to be product states, but when Lemma 5 was used to get Eq.(10) in page 110, the states $\gamma_{j,t}$ are not product states (generally, they can be entangled states). Thus, the procedure to get Eq.(10) may have a bug, whereas one can find that Eq.(10) is a crucial step, without which the lower bound may not be obtained for CL-1QFA.
In the following, we present another method which  can not only fix the possible bug in \cite{BMP14}, but also  deal with uniformly lower bounds on the size of  1QCFA, CL-1QFA and 1QFAC.

Before that, we first recall some preliminary knowledge. A language $L$ is said to be recognized by  QFA ${\cal A}$ with cut-point $\lambda\in(0,1]$, if ${\cal P}_{\cal A}(x)>\lambda$ holds for all $x\in L$ and
${\cal P}_{\cal A}(x)\leq\lambda$ holds for all $x\notin L$. The cut-point $\lambda$ is said to be {\it isolated} whenever there exists $\delta\in(0, \frac{1}{2}]$ such that $|{\cal P}_{\cal A}(x)-\lambda|\geq \delta$, for all $x\in\Sigma^*$. In this case,  we simply say that $L$ is recognized by ${\cal A}$ with  cut-point isolated by $\delta$.

Two QFA  ${\cal A}_1$ and  ${\cal A}_2$ over $\Sigma$ are said to be equivalent, if ${\cal P}_{{\cal A}_1}(x)={\cal P}_{{\cal A}_2}(x)$ holds for all $x\in \Sigma^*$.

Below we first recall a result  given in \cite{LF14} that is useful for obtaining our results in this paper.

\begin{thm}
For any semi-quantum automaton (including CL-1QFA, 1QFAC and 1QCFA) with $q$ quantum basis states and $k$ classical states, there is a $kq$-dimensional MO-1gQFA equivalent to it. \label{thm-simulation}
\end{thm}

{\it Proof.}   For readability,  we present here the proof for CL-1QFA and one can refer to \cite{LF14} for other cases.
Let  CL-1QFA  ${\cal
  A}=(Q, \Sigma, {\cal C},  q_1, \{U_\sigma\}_{\sigma\in\Sigma}, \mathcal{M}, {\cal L})$,  with ${\cal L}$ accepted by DFA $\mathcal{A}'=(S,{\cal C}, s_1,\delta,  S_a)$.  We construct an MO-1gQFA $$\widehat{\cal A}=({\cal H},\Sigma, \rho_0, \{{\mathcal{E}}_\sigma\}_{\sigma\in\Sigma}, P_{acc})$$ from $\mathcal{A}$ and $\mathcal{A}'$ as follows:
\begin{itemize}
  \item  ${\cal H}={\cal H}_Q\otimes{\cal H}_S$ where ${\cal H}_A$ denotes the Hilbert space spanned by $A$;
   \item $ \rho_0=|q_1\rangle \langle q_1|\otimes |s_1\rangle \langle s_1|$;
\item ${P}_{acc}=I_Q\otimes\sum_{s\in S_a} |s\rangle\langle s|$ where $I_Q$ is the identity operator on ${\cal H}_Q$;
  \item for each $\sigma\in\Sigma$, ${\mathcal{E}}_{\sigma}$  has operation elements $\{ {E}_\sigma^{c,s}\}_{c\in \mathcal{C}, s\in S}$ where
      \begin{align} {E}_\sigma^{c,s}=P_cU_\sigma\otimes |\delta(s,c)\rangle \langle s|.\end{align}
\end{itemize}

 It is easy to verify that the  collection of operators $\{ {E}_\sigma^{c,s}\}$ satisfies the completeness condition.
Furthermore, for $\rho\otimes \varrho\in L(\mathcal{H}_Q\otimes \mathcal{H}_S)$, we have
\begin{align}{\mathcal{E}}_\sigma(\rho\otimes \varrho)=\sum_{c\in \mathcal{C}}P_cU_\sigma\rho U_\sigma^\dagger P_c\otimes \mathcal{F}_c(\varrho),\end{align}
where $\mathcal{F}_c(\varrho)=\sum_{s\in S}|\delta(s,c)\rangle\langle s|\varrho|s\rangle\langle \delta(s,c)|$.

Now let us check the behavior of $\widehat{\cal A}$ on an input string. Suppose  $\widehat{\cal A}$ starts with the initial state $\rho_0$ and scans a symbol $\sigma$. Then the resulting state is
\begin{align*}\rho&={\mathcal{E}}_\sigma(|q_1\rangle \langle q_1|\otimes |s_1\rangle \langle s_1|)\\
&=\sum_{c\in{\cal C}}P_{c}U_\sigma|q_1\rangle \langle q_1|U_\sigma^\dagger P_{c}\otimes\mathcal{ F}_{c}(|s_1\rangle\langle s_1|)\\
&=\sum_{c\in{\cal C}}P_{c}U_\sigma|q_1\rangle \langle q_1|U_\sigma^\dagger P_{c}\otimes |t_c\rangle\langle t_c|,\end{align*}
where $t_c=\delta(s_1,c)$. In this way, after scanning a string $x=x_1x_2\cdots x_n\in\Sigma^*$,  the final state is
\begin{align}\rho_x=\sum_{y\in{\cal C}^n}|\phi_y\rangle \langle\phi_y|\otimes |s_y\rangle\langle s_y|,\end{align}
where  $|\phi_y\rangle=\prod_{i=1}^n(P_{y_i}U_{x_i})|q_1\rangle$ and $s_y=\delta^*(s_1,y)$.
 Note that $s_y\in S_a$ iff $y\in {\cal L}$.  Thus the  probability of $\widehat{\cal A}$ accepting $x$ is
\begin{align*}P_{\widehat{\cal A}}(x)&=\text{Tr}(\widehat{P}_a\rho_x)=\sum_{y\in{\cal C}^n}\chi_a(s_{y})\||\phi_y\rangle \|^2 \\
&=\sum_{y_1y_2\cdots y_n\in{\cal L}}\left\|\prod^{n}_{i=1}(P_{y_i}U_{x_i})|q_1\rangle\right\|^2,
 \end{align*}
where $\chi_a(s)$ is $1$ if $s$ is in $S_a$ and $0$ else. Note that the above probability is equal to the one of ${\cal A}$ given in Eq. (\ref{f_CL}).
Therefore, we have completed the proof.\qed

Now we are in a position to give our main results.
\begin{thm}
Suppose that $L$ is recognized  by an $n$-dimensional MO-1gQFA with cut-point isolated by $\delta$. Then $L$ can be recognized by  a DFA with $d$ states satisfying
\begin{align}d\leq\left(1+\frac{\sqrt{n}}{\delta}\right)^{2n^2}.\label{UBD}\end{align}
\end{thm}

{\noindent \it Proof.}  Assume that $L$ is  recognized by
 MO-1gQFA ${\cal M}=\{ {\cal H},\Sigma,\rho_0,\{{\cal
E}_\sigma\}_{\sigma\in\Sigma}, P_{acc}\}$  with cut-point isolated by $\delta$. An equivalence
relation ``$\equiv_L$'' on $x,y\in\Sigma^*$ is defined by: $x\equiv_L y$ if
for all $z\in\Sigma^*$, $xz\in L$ iff $yz\in
L$. Then in terms of Myhill-Nerode theorem \cite{HU79}, it  suffices to prove that   the number of
equivalence classes induced by ``$\equiv_L$'' is upper bounded by the right side of (\ref{UBD}).

Let $\rho_x={\cal E}_{x_n}\circ\dots\circ{\cal E}_{x_2}\circ{\cal
E}_{x_1}(\rho_0)$, i.e.,  the state of ${\cal M}$ after scanning the word $x$.  Now, suppose that $x\not\equiv_L y$, that is,
there exists a string $z\in\Sigma^*$ such that $xz\in L$ and
$yz\notin L$. Then we have
\begin{align}
\text{Tr}(P_{acc}{\cal
E}_z(\rho_x))\geq\lambda+\delta~~\text{and}~~\text{Tr}(P_{acc}{\cal
E}_z(\rho_y))\leq\lambda-\delta
\end{align}
for some $\lambda\in(0,1]$, where ${\cal E}_z$ stands for ${\cal
E}_{z_m}\circ\dots\circ{\cal E}_{z_2}\circ{\cal E}_{z_1}$.  Denote
\begin{align*} &p_{acc}=\text{Tr}(P_{acc}{\cal E}_z(\rho_x)), && p_{rej}=\text{Tr}(P_{rej}{\cal E}_z(\rho_x)),\\
&q_{acc}=\text{Tr}(P_{acc}{\cal E}_z(\rho_y)), && q_{rej}=\text{Tr}(P_{rej}{\cal E}_z(\rho_y)).
\end{align*}
Then we have
\begin{align*}
\sqrt{n}||vec(\rho_x)-vec(\rho_y)||&\geq||\rho_x-\rho_y||_{tr} &&~~~\text{(by Lemma \ref{lm-norm})}\\
&\geq||{\cal E}_z(\rho_x)-{\cal
E}_z(\rho_y)||_{tr}  &&~~~\text{(by (i) of Lemma \ref{lm-distance})} \\
&\geq|p_{acc}-q_{acc}|+|p_{rej}-q_{rej}| &&~~~\text{(by (ii) of Lemma \ref{lm-distance})} \\
&\geq2\delta
\end{align*}
On the other hand, for any $x\in \Sigma^*$, we have
\begin{align}||vec(\rho_x)||\leq \|\rho_x\|_{tr}=\text{Tr}(\rho_x)=\text{Tr}(\rho_0)=1\end{align}
where the inequality follows from Lemma \ref{lm-norm}, the  first equality holds because $\rho_x$ is positive semi-definite, and the second equality holds because the operations used are trace-preserving.

In summary, we obtain the following two properties:
\begin{itemize}
  \item For any $x\in \Sigma^*$, $vec(\rho_x)$ lies in the unit sphere in $\mathbb{C}^{n^2}$.
  \item For any two  strings $x,y\in\Sigma^*$ satisfying
$x\not\equiv_L y$, we always have
\begin{align}
||vec(\rho_x)-vec(\rho_y)||_{tr}\geq\frac{2\delta}{\sqrt{n}}.\label{X-Y}
\end{align}
\end{itemize}

Now, suppose that $\Sigma^{*}$ consists of $d$ equivalence
classes, say $[x_1],[x_2],\cdots, [x_d]$. Arbitrarily choose an element $vec(\rho_{x_i})$. Let\begin{align}U\left(vec(\rho_{x_i}), \frac{\delta}{\sqrt{n}}\right)=\left\{|\phi\rangle: \||\phi\rangle-vec(\rho_{x_i})\|\leq\frac{\delta}{\sqrt{n}}\right\},\end{align} i.e., a sphere centered at $vec(\rho_{x_i})$ with the radius $\frac{\delta}{\sqrt{n}}$. Then  all these spheres do not intersect pairwise except for  their surface, and all of them are contained in a large sphere in $\mathbb{C}^{n^2}$ centered at $(0,0,\cdots,0)$ with the radius $1+\frac{\delta}{\sqrt{n}}$.  Note that $\mathbb{C}^{n^2}$ is an $n^2$-dimensional complex space and each element from it can be represented by an element of $\mathbb{R}^{2n^2}$. Then the volume of a sphere of a radius $r$ in $\mathbb{C}^{n^2}$ is $cr^{2n^2}$ where $c$ depends only on $n$. Therefore, it holds that
\begin{equation}
d\leq\frac{c(1+\frac{\delta}{\sqrt{n}})^{2n^2}}{c(\frac{\delta}{\sqrt{n}})^{2n^2}}=(1+\frac{\sqrt{n}}{\delta})^{2n^2}.
\end{equation}
This completes the proof.\qed\\

Furthermore, we have
\begin{thm}
 Let $L$ be a regular language  whose minimal DFA has $d$ states. Then any $n$-dimensional MO-1gQFA recognizing $L$ with cut-point isolated by $\delta$  must satisfy
\begin{align}n\geq \left[\frac{\log d}{2\log\frac{2}{\delta}}\right]^{\frac{4}{9}}.\end{align}
\end{thm}
{\noindent \it Proof.} For $n\geq1$ and $\delta\in(0,\frac{1}{2}]$, we have \begin{align}d\leq\left(1+\frac{\sqrt{n}}{\delta}\right)^{2n^2}\leq\left(\frac{2}{\delta}\right)^{2\sqrt[4]{n}n^2}= \left(\frac{2}{\delta}\right)^{
2n^{\frac{9}{4}}},\end{align}
where the second inequality holds because $1+\frac{\sqrt{x}}{\delta}\leq \left(\frac{2}{\delta}\right)^{x^{\frac{1}{4}}}$ holds for any $x\geq1$ and $\delta\in(0,\frac{1}{2}]$.\qed\\

Therefore, based on Theorem \ref{thm-simulation} and the above theorem, we obtain a lower bound on the size of semi-quantum automata as follows.
\begin{thm}
Suppose a semi-quantum automaton (including CL-1QFA, 1QFAC and 1QCFA), with $q$ quantum basis states and $k$ classical states, recognizes  a regular language whose minimal DFA has $d$ states, with cut-point isolated by $\delta$. Then it holds that
\begin{align}qk\geq \left[\frac{\log d}{2\log\frac{2}{\delta}}\right]^{\frac{4}{9}}.\end{align}
\end{thm}

\begin{Rm}Not that recently Ref. \cite{BMP14} dealt with the size lower bound of CL-1QFA. Compared with that, our work has the following advantages: (i) our method is much more concise,  whereas   complex technical treatments were used in \cite{BMP14}; and (ii) our method is universal, since it is applicable to the three existing models of semi-quantum automata, while the method in  \cite{BMP14} was tailored for CL-1QFA. In addition, our bound is slightly more optimal than the one in \cite{BMP14}, since the lower bound  obtained in \cite{BMP14} actually should be $\left[\frac{\log d}{2\log\frac{5}{\delta}}\right]^{\frac{4}{9}}$, although it was claimed to be $\left[\frac{\log d}{\log\frac{5}{\delta}}\right]^{\frac{4}{9}}$ (the factor $2$ is from the fact the volume of a sphere of a radius $r$ in $\mathbb{C}^{n^2}$ is $cr^{2n^2}$, instead of $cr^{n^2}$, where $c$ depends only on $n$).
\end{Rm}
\section{Conclusions}
We have presented a uniform method for obtaining the lower bound on the size of CL-1QFA, 1QFAC and 1QCFA, and this bound shows that these automata can be at most exponentially smaller than DFA. Compared with a recent work \cite{BMP14}, our method is much more concise and universal, and it is applicable  to the three existing main models of semi-quantum automata.

Note that although  our lower bound is universal, it is not necessarily optimal. For instance, a better lower bound $\log d/2\log(1+\frac{\sqrt{2}}{\delta})$ was obtained for 1QFAC in \cite{QMS09}. Thus, a natural open problem remains either to witness the optimality of our size lower bound for some specific model, or to improve it.

\section*{Acknowledgements}
The authors are thankful to Dr. Shenggen Zheng for his useful comments.

\end{document}